\documentclass[apjl]{emulateapj}

\newcommand{\kms}{\,km\,s$^{-1}$}     
\newcommand{\ha}{\,H$\alpha$}     

\newcommand{\arcs}{$^{\prime\prime}$}

\newcommand{\fuse}{$FUSE$}

\journalinfo{To be published in ApJ Letters} 
\shorttitle{O VI in the M82 Starburst Superwind}
\shortauthors{Hoopes et al.}

\begin{document}

\title{Cooling in Coronal Gas in the M82 Starburst Superwind\altaffilmark{1}}

\author{Charles G. Hoopes\altaffilmark{2}, Timothy M. Heckman\altaffilmark{2}, David K. Strickland\altaffilmark{2}, and J. Christopher Howk\altaffilmark{3}} 
\altaffiltext{1}{Based on observations made with the NASA-CNES-CSA Far Ultraviolet Spectroscopic Explorer. FUSE is operated for NASA by the Johns Hopkins University under NASA contract NAS5-32985.}
\altaffiltext{2}{Department of Physics and Astronomy, Johns Hopkins University,
3400 N. Charles St., Baltimore, MD 21218; choopes@pha.jhu.edu, heckman@pha.jhu.edu, dks@pha.jhu.edu}
\altaffiltext{3}{Center for Astrophysics and Space Science, University of California at San Diego, C-0424, La Jolla, CA 92093; howk@ucsd.edu}

\begin{abstract}

We have used the {\it Far Ultraviolet Spectroscopic Explorer} to
search for \ion{O}{6} $\lambda1031.926$ emission at
four locations in the starburst superwind of M82.  No \ion{O}{6}
emission was detected at any of the four pointings, with upper limits
less than or equal to the 0.3$-$2 keV X-ray flux. These observations
limit the energy lost through radiative cooling of coronal phase
($T\sim10^{5.5}$~K) gas to roughly the same magnitude as that lost in
the hot phase through X-ray emission, which has been shown to be
small. The wind material retains most of its energy and should be able
to escape from the gravitational potential of M82, enriching
the intergalactic medium with energy and metals. The lack of coronal
gas in the wind and observations of spatially correlated X-ray and
\ha\ emission are consistent with a scenario in which the hot wind
material over-runs cold clouds in the halo, or one where the \ha\ and
X-ray emission arise at the interface between the hot wind and a cool
shell of swept-up ISM, as long as the shock velocity is
$\la150$~\kms. The observed limits on the \ion{O}{6}/H$\alpha$ and
\ion{C}{3}/H$\alpha$ flux ratios rule out shock heating as the source of the
$T=10^4$~K gas unless the shock velocity is $\la90$~\kms.

\end{abstract}


\section{Introduction}

A common product of starbursts in galaxies is a galactic-scale outflow
of gas and energy from the starburst region, called a starburst
superwind (Heckman, Armus, \& Miley 1990; Strickland \& Stevens
2000). A superwind begins as an expanding bubble of hot gas powered by
the combined energy of stellar winds and supernovae. This bubble may
break out of the disk, resulting in a wind of hot ($T\sim10^{8}$~K),
low-density gas flowing into the halo \citep{cc85}. As this hot gas
collides with cold ambient material in the halo, intermediate
temperature gas ($T\sim10^{6}-10^{7}$~K) is formed at the
interface. This gas emits X-rays, and if it cools through
$T\sim10^{5.5}$~K it will produce \ion{O}{6} $\lambda\lambda1031.926,
1037.617$ line emission.

The question of whether superwinds are able to overcome the
gravitational potential of their host galaxies is critical for
determining their full impact on the evolution of galaxies and the
intergalactic medium (IGM). If the wind does escape, the wind material
will be deposited into the IGM, enriching it with energy and metals
({\it e.g.,} Aguirre et al. 2001), and it may produce the observed
mass-metallicity relationship in galaxies if winds are preferentially
able to escape from low mass systems \citep{m99, hlsa00}. A comparison
of wind velocities with escape velocities suggests that winds can
readily escape from dwarf galaxies, while winds of larger galaxies
cannot \citep{hlsa00}. However, wind velocities have only been
measured relatively close to the starburst region, so it is still
unclear whether energy loss further downstream keeps the material
bound to the galaxy. Observations and models imply that radiative
cooling through X-ray emission from hot gas is not effective at
removing kinetic energy from superwinds
\citep{dks00}. However, coronal temperature gas at $T\sim10^{5.5}$~K
cools very quickly and efficiently \citep{sd93}, so energy lost
through this phase may be enough to prevent winds from escaping.

The \ion{O}{6} doublet at 1031.926, 1037.617~\AA\ is the most
important coolant for coronal gas \citep{edgar86}, making
\ion{O}{6} emission an excellent measure of the energy lost in the
coronal phase. This wavelength range is inaccessible from ground-based
telescopes, and opportunities to observe it from space have been
limited. This has changed with the launch of {\it Far Ultraviolet
Spectroscopic Explorer} (\fuse; Moos et al. 2000). We have used \fuse\
to search for \ion{O}{6} emission in the halo of M82, the prototypical
starburst superwind galaxy.

\section{Observations and Data Reduction}

\begin{figure*}
\epsscale{1.0}
\plottwo{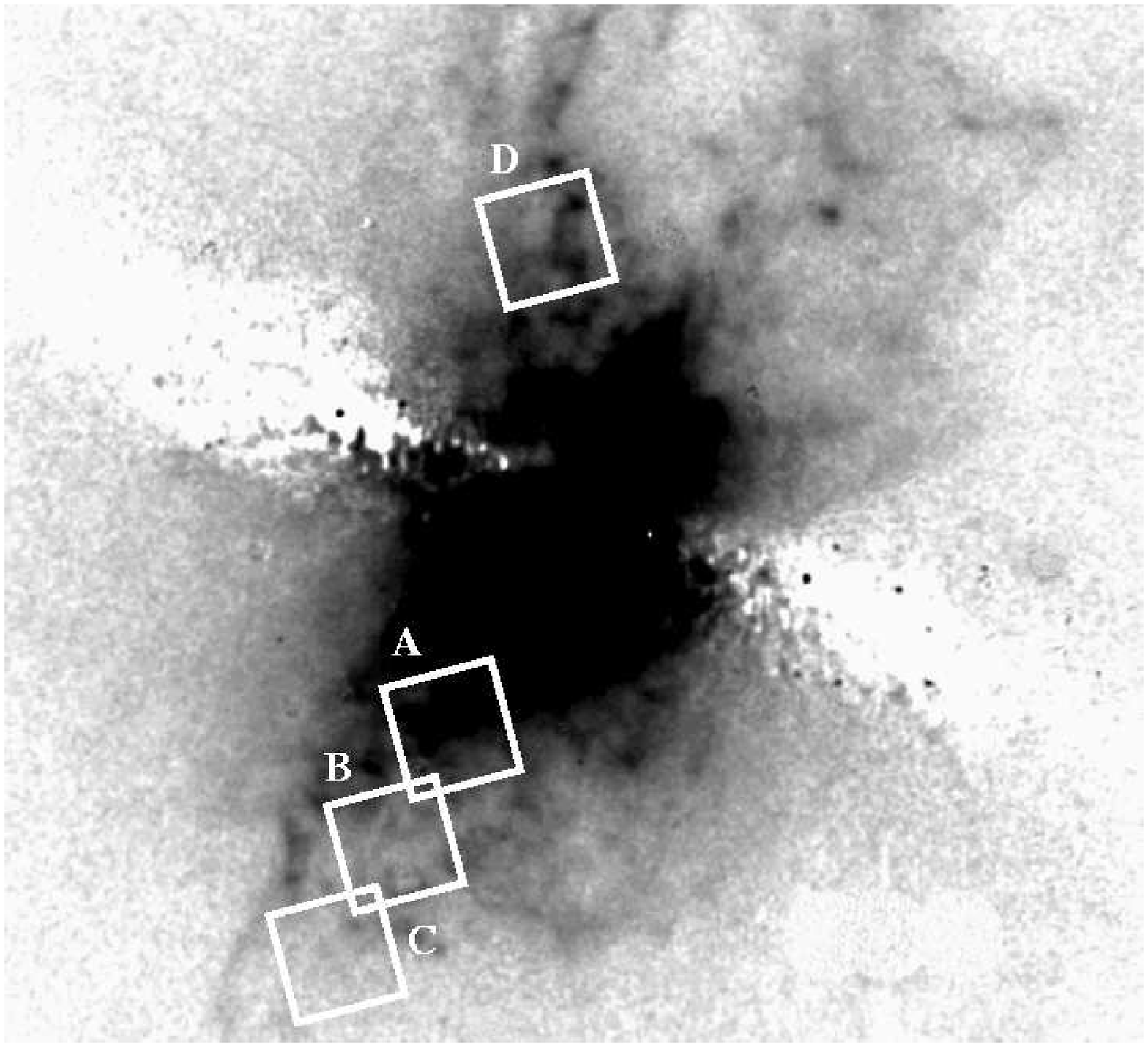}{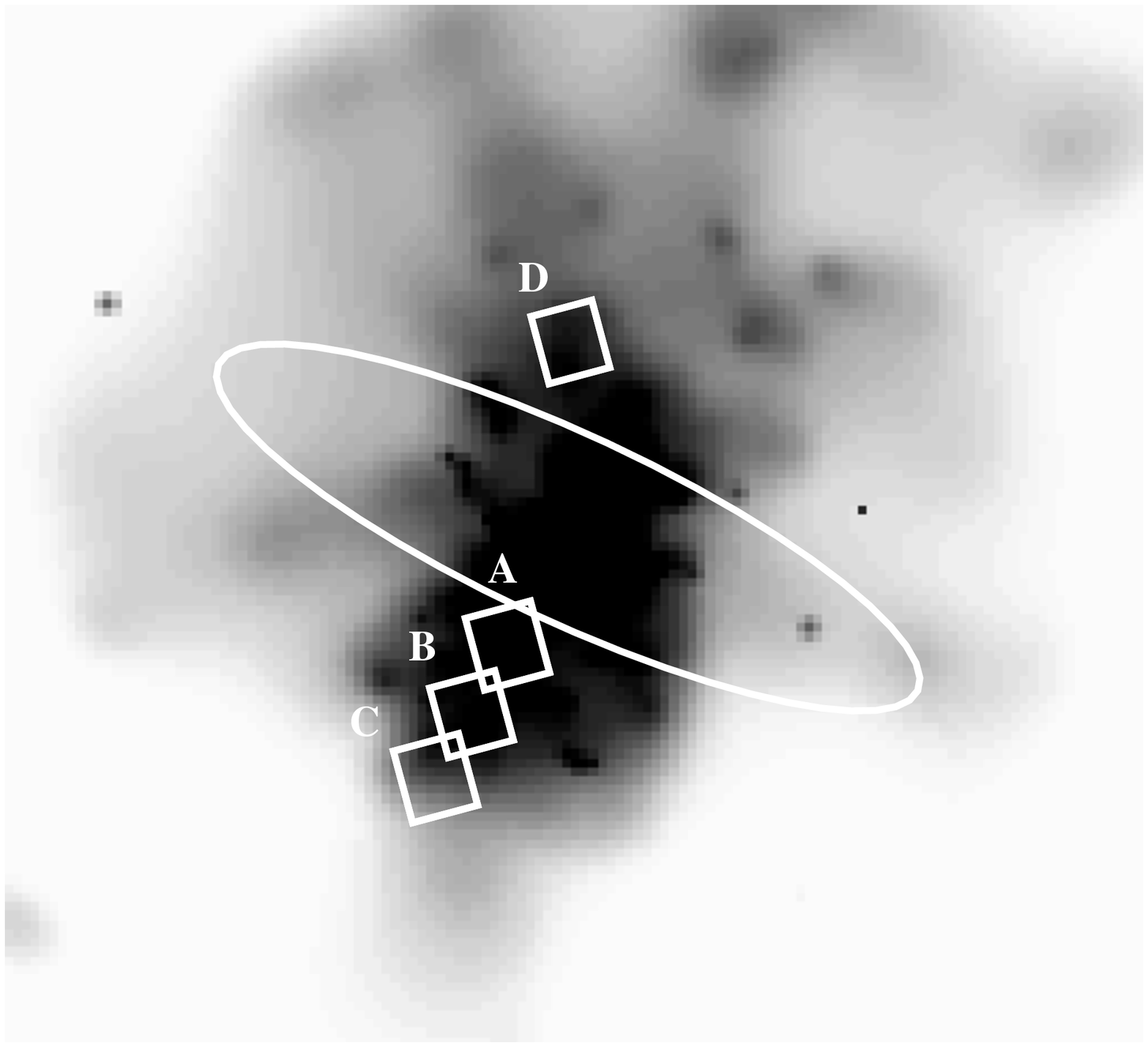}
\caption{{\it Left}: An H$\alpha$ image of M82 \citep{c96}, showing the
locations of the {\em FUSE} pointings. {\it Right}: The {\em Chandra}
$0.3-2.0$~keV image of M82 (Strickland et al. 2003). The ellipse shows the
approximate location of the optical disk of M82.}
\end{figure*}

We obtained \fuse\ spectra at four positions in the halo of M82 on
2002 February 3. The coordinates and exposure times are listed in
Table 1, and the positions of the pointings can be seen in Figure
1. The large (LWRS: $30$\arcs$\times30$\arcs) apertures were
used. \fuse\ consists of four separate telescopes, two with LiF coated
optics for optimal sensitivity to longer wavelengths (1000 to
1180~\AA) and two with SiC coated optics for optimal sensitivity to
shorter wavelengths (900 to 1000~\AA). These are usually referred to
as the LiF1, LiF2, SiC1, and SiC2 channels. Of these, the LiF channels
have the largest effective area at 1032~\AA.

The raw photon event lists were screened to remove data taken while
the satellite was in the South Atlantic Anomaly, and to remove events
with pulse height amplitudes less than 4 and greater than 15, because
these are predominantly background events. This photon list was then
transformed to a 2-dimensional image from which we extracted the
spectrum, using bright airglow lines to determine its location on the
detector. The extracted spectra from the individual exposures were
co-added to produce the final spectrum for each pointing. The
wavelength solution from the CALFUSE pipeline (v2.0.5) was applied to
the data, but we kept the data in units of counts. The individual
channels were not co-added.

\begin{deluxetable}{lccc}[hb!]
\tablewidth{0pc}
\tablecaption{\fuse\ Observations} 
\tablehead{\colhead{Pointing} & \colhead{R.A.} & \colhead{Decl.} & \colhead{Exposure Time}\\
\colhead{} & \colhead{(J2000.0)}  & \colhead{(J2000.0)} & \colhead{(s)} }
\startdata
A    & 9~55~57.1 & 69~39~49.5 & 8056  \\
B    & 9~55~60.0 & 69~39~17.1 & 8768  \\
C    & 9~56~03.1 & 69~38~48.0 & 11529  \\
D    & 9~55~52.3 & 69~42~54.5 & 10011  \\
\enddata
\end{deluxetable}

\section{Limits on O VI Emission}

No \ion{O}{6} $\lambda1031.926$ or $\lambda1037.617$ emission was
evident in any of the spectra. We also searched for \ion{C}{3}
$\lambda977.020$ emission in the SiC channels, finding none. We base
our \ion{O}{6} limits on the 1031.926~\AA\ line because it is the
strongest of the \ion{O}{6} doublet. The background flux level was
determined by taking the mean of the flux measured in ten 0.34~\AA\
bins near 1032~\AA.  The noise in the background is the
primary source of uncertainty in the \ion{O}{6} measurement, so the
square root of the background flux corresponds to the $1\sigma$
uncertainty, from which we derive $3\sigma$ \ion{O}{6} detection
limits of the FUSE spectra.

The choice of 0.34~\AA\ bins for computing the sensitivity corresponds
to the filled aperture resolution for the LWRS aperture, and also
matches the expected velocity range of \ion{O}{6} emission based on
absorption observations of NGC~1705 \citep{heckman01}. However,
\cite{sb98} found that the \ha\ emission is spread over $300$~\kms\ in
the M82 outflow, which would correspond to an emission FWHM of 1~\AA\
at 1032~\AA. Using this bin size would increase the observed
\ion{O}{6} upper limits by a factor of 1.7. We adopt the former value
(FWHM$=100$~\kms) since it is based on an \ion{O}{6} measurement, but
the difference is not enough to change our conclusions.

These limits were converted from counts to flux using the effective
area of the LiF1 channel at 1032~\AA\ and the exposure time. The
observed limit was corrected for Galactic extinction of A$_V$=0.10
\citep{bh84} and extinction intrinsic to the M82 superwind A$_V=0.85$
\citep{ham90}. We also estimated the extinction by modeling the
photoelectric absorption in {\em Chandra} $0.3-2.0$~keV X-ray data
\citep{dks03}, and found agreement with the
extinction derived from the Balmer decrement by \cite{ham90}. The
X-ray opacity is dependent upon dust grains and the gas-phase metal
column, so it is directly related to the ISM constituents responsible
for the far-ultraviolet extinction.  We corrected the internal
extinction three different ways: using the Galactic extinction law
\citep{cardelli89}, the LMC extinction law \citep{howarth83}, and the
starburst extinction law
\citep{calzetti94}. The LMC law produces more far-ultraviolet extinction than
the Galactic law, while the starburst law produces much less. We adopt
the Galactic extinction law as the intermediate case, although using
any of the three laws would not alter our main conclusions. The
extinction-corrected limits are listed in Table 2. Figure 2 shows the
observed spectrum of position A (see Figure 1), along with a model
of the expected \ion{O}{6} emission (if it were equal to the X-ray
flux).

\begin{deluxetable*}{lcccccc}
\tablewidth{0pc}
\tabletypesize{\scriptsize}
\tablecaption{Measured Properties\tablenotemark{a}} 
\tablehead{\colhead{Pointing} & 
\colhead{F$_{1032}$ Limit\tablenotemark{b}} &
\colhead{F$_{coronal}$ Limit\tablenotemark{c}} & \colhead{F$_{coronal}$/F$_{X-ray}$\tablenotemark{d}} & \colhead{F$_{C~III}$ Limit\tablenotemark{b}} &
\colhead{F$_{O~VI}$/F$_{H\alpha}$\tablenotemark{e}} &\colhead{F$_{C~III}$/F$_{H\alpha}$\tablenotemark{e}}\\
\colhead{} & \colhead{(erg cm$^{-2}$ s$^{-1}$)}& \colhead{(erg cm$^{-2}$ s$^{-1}$)}& \colhead{(erg cm$^{-2}$ s$^{-1}$)}& 
\colhead{} &\colhead{} &\colhead{} }
\startdata
A    & $<7.6\times10^{-14}$ & $<3.8\times10^{-13}$ & $<0.40$ & $<4.4\times10^{-13}$ & $<0.01$ & $<0.05$\\
B    & $<7.6\times10^{-14}$ & $<3.8\times10^{-13}$ & $<0.54$ & $<4.4\times10^{-13}$ & $<0.03$ & $<0.18$\\
C    & $<6.2\times10^{-14}$ & $<3.1\times10^{-13}$ & $<2.58$ & $<3.2\times10^{-13}$ & $<0.06$ & $<0.29$\\
D    & $<6.3\times10^{-14}$ & $<3.2\times10^{-13}$ & $<5.08$ & $<3.5\times10^{-13}$ & $<0.02$ & $<0.09$\\
\enddata
\tablenotetext{a}{Upper limits are $3\sigma$.}
\tablenotetext{b}{Corrected for extinction, assuming a Galactic
extinction of A$_V$=0.10 magnitudes, and an M82 extinction of
A$_V$=0.85 (Heckman et al. 1990), and using the Galactic extinction
law (Cardelli et al. 1989), which gives $A_{1032}=4.84A_V$ (assuming $R_V$=3.1). The LMC extinction law (Howarth 1983) gives $A_{1032}=5.55A_V$, which would result in upper limits $\sim1.8\times$ greater, and the starburst extinction law (Calzetti et al. 1994) gives $A_{1032}=2.35A_V$ (normalized to match the Galactic law at 5500\AA), which would result in upper limits $\sim7.0\times$ lower.}
\tablenotetext{c}{Total energy radiated by the coronal gas, corrected for extinction.}
\tablenotetext{d}{The absorption corrected X-ray flux (0.3-2.0 keV) within the $30^{\prime\prime}\times30^{\prime\prime}$ LWRS
aperture was measured on Chandra ACIS-S spectra (Strickland et al. 2003).}
\tablenotetext{e}{The extinction corrected H$\alpha$ flux was measured in the $30^{\prime\prime}\times30^{\prime\prime}$ LWRS
aperture using the image shown in Figure 1.}
\end{deluxetable*}

Position D is on the North side of the M82 outflow, and since the
sight line passes through the disk of M82 it is likely subject to
higher extinction. The optical spectrum of \cite{ham90} did not
reach this far into the halo, so we have no measure of the extinction
in this region. The most reliable limits come from positions A, B, and
C, which are on the southwest (near) side of the outflow and do not
pass through the M82 disk.

The velocity of M82 ($v_r=203$~\kms, de Vaucouleurs et al. 1991)
shifts the \ion{O}{6} emission to 1032.624~\AA. The only Galactic
absorption nearby is the R(4)~$6-0$ line of H$_2$ at 1032.351~\AA. An
examination of sight lines near M82 ($\sim7^{\circ}$ away) observed by
\fuse\ suggests that H$_2$ in the $J=4$ rotational level is present in
this direction. This line is $\sim80$~\kms\ from the expected center
of the \ion{O}{6} line, so the contamination is expected to be minimal
even if the lines are broad. The M82 \ion{C}{3} line falls at
977.681~\AA, which is in a region free of Galactic absorption, but is
85~\kms\ from an atmospheric \ion{O}{1} emission line at 977.959~\AA\
\citep{f01}.

In the optically thin case, the energy radiated through the two
\ion{O}{6} lines is 150\% of the 1032~\AA\ line strenth. The two
\ion{O}{6} lines together are responsible for $\sim30$\% of the cooling
in coronal gas \citep{heckman01}. The limit on the total energy
radiated by coronal gas $F_{coronal}$ is thus a factor of 5 higher
than the limit on 1032~\AA\ flux (see Table 2). These limits are
comparable to the observed X-ray flux at each position, ranging from
40\% to 260\% in positions A, B, and C. \cite{dks00} found that the
energy lost through X-ray emission in M82 is $\le10$\% of the energy
input from the starburst. If the lowest upper limit on \ion{O}{6} is
valid everywhere in the wind, then the amount of energy radiated away
by the coronal gas is even smaller than that lost through X-rays.

\section{Discussion}

\subsection{Radiative Cooling in the Superwind}

The non-detection of \ion{O}{6} emission has profound implications for
the evolution of the starburst superwind in M82. The lack of
\ion{O}{6} emission indicates that radiative cooling from coronal gas
does not effectively remove energy from the wind. Energy loss through
other phases has been determined to be insignificant as well
\citep{dks00}, so the wind must retain most of its kinetic/thermal
energy as it travels through the halo. The wind velocity comfortably
exceeds the escape velocity, so the wind should easily escape from the
galaxy.

\cite{o03} detected \ion{O}{6} emission in the halo of NGC~4631 at the
locations of strong X-ray emission. NGC~4631 is actively forming stars
but is not a starburst and has weaker X-ray emission overall than
M82. Nevertheless, the \ion{O}{6}/X-ray ratios were $\sim1$, similar
to our limits in M82, suggesting that radiative cooling of coronal gas
is ineffective in the hot halo of NGC~4631, a conclusion we also reach
for the M82 superwind. \cite{heckman01} measured \ion{O}{6} absorption
and set upper bounds on the \ion{O}{6} emission in the starburst
NGC~1705, and were able to limit the cooling rate to $\le20$\% of the
supernova heating rate in that galaxy. Since the X-ray luminosity of
NGC~1705 is very low, the NGC~1705 superwind should also escape into
the IGM. If these results hold for many starbursts, they could be an
important source of energy and metals for the IGM.

\subsection{Origin of \ion{O}{6}, \ha, and X-rays}

\begin{figure}[b!]
\epsscale{1.25}
\plotone{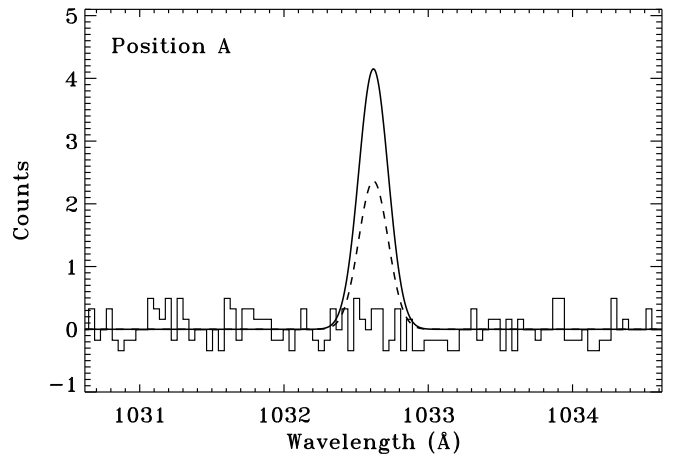}
\caption{The {\em FUSE} spectrum
of the wavelength region around 1032~\AA\ in position A. The observed
spectrum (histogram) is binned to 0.037~\AA/pixel, and a constant
background was subtracted. No O VI emission is visible in the
spectrum. The smooth lines show the expected number of counts if the
O~VI emission were equal to the observed 0.3-2.0 keV X-ray flux, after
accounting for extinction using the Galactic extinction law (Cardelli
et al. 1989, solid line), and the LMC extinction law (Howarth et
al. 1983, dashed line). The FWHM of the emission line is assumed to be
0.34~\AA\ (100~km~s$^{-1}$).}
\end{figure}

\cite{dks02} found significant spatial correlation between \ha\ and
X-ray emission in the starburst superwind of NGC~253, and a similar
correlation is seen in the M82 superwind \citep{l99,dks03}. This fact
strongly suggests a physical connection between the origin of both
types of emission. Models of superwinds must account for this
correlation, and \cite{dks02} outline several possibilities. The
\fuse\ pointings were chosen because they exhibit strong X-ray
emission and \ha\ emission, suggesting the presence of warm and hot
gas (see Figure 1).

It is unlikely that both the X-ray and \ha\ emission arise in cooling
wind material. \cite{dks02} showed that the cooling times are too
long, and the non-detection of \ion{O}{6} emission indicates that
significant amounts of gas do not cool through $T=10^{5.5}$~K. Another
possibility is that the \ha\ emission arises in a cool shell of swept
up ISM surrounding the superwind ({\it e.g.} Weaver et al. 1977). If
the shock velocity $v_s\ga150$~\kms\ the interface between the hot wind
fluid and the cool shell material would contain coronal gas, and
\cite{dks02} showed that the luminosity of the \ion{O}{6} doublet would
be $\sim100$ times greater than the X-ray luminosity. This is clearly
at odds with the observed \ion{O}{6} upper limits, but we cannot rule
out a lower $v_s$. We also cannot rule out a {\it hot} shell of
swept-up ISM which would not produce coronal gas due to long cooling
times, but this scenario has difficulty reproducing the spatial
correlation between the X-ray and \ha\ emission
\citep{dks02}.

\cite{l99} discuss a model in which the wind runs into cool clouds in
the halo, either pre-existing or entrained within the wind itself (see
also Strickland et al. 2002).  The X-ray emission is produced in a
stand-off bow shock upstream from a cloud, while the \ha\ is produced by a
slow shock driven into the cloud by the wind (with a possible
contribution from photoionization).  The X-ray and \ha\ emission in
this scenario are physically correlated, but little coronal
temperature gas is produced, consistent with our non-detection of
\ion{O}{6} emission.

\begin{figure}
\epsscale{1.25}
\plotone{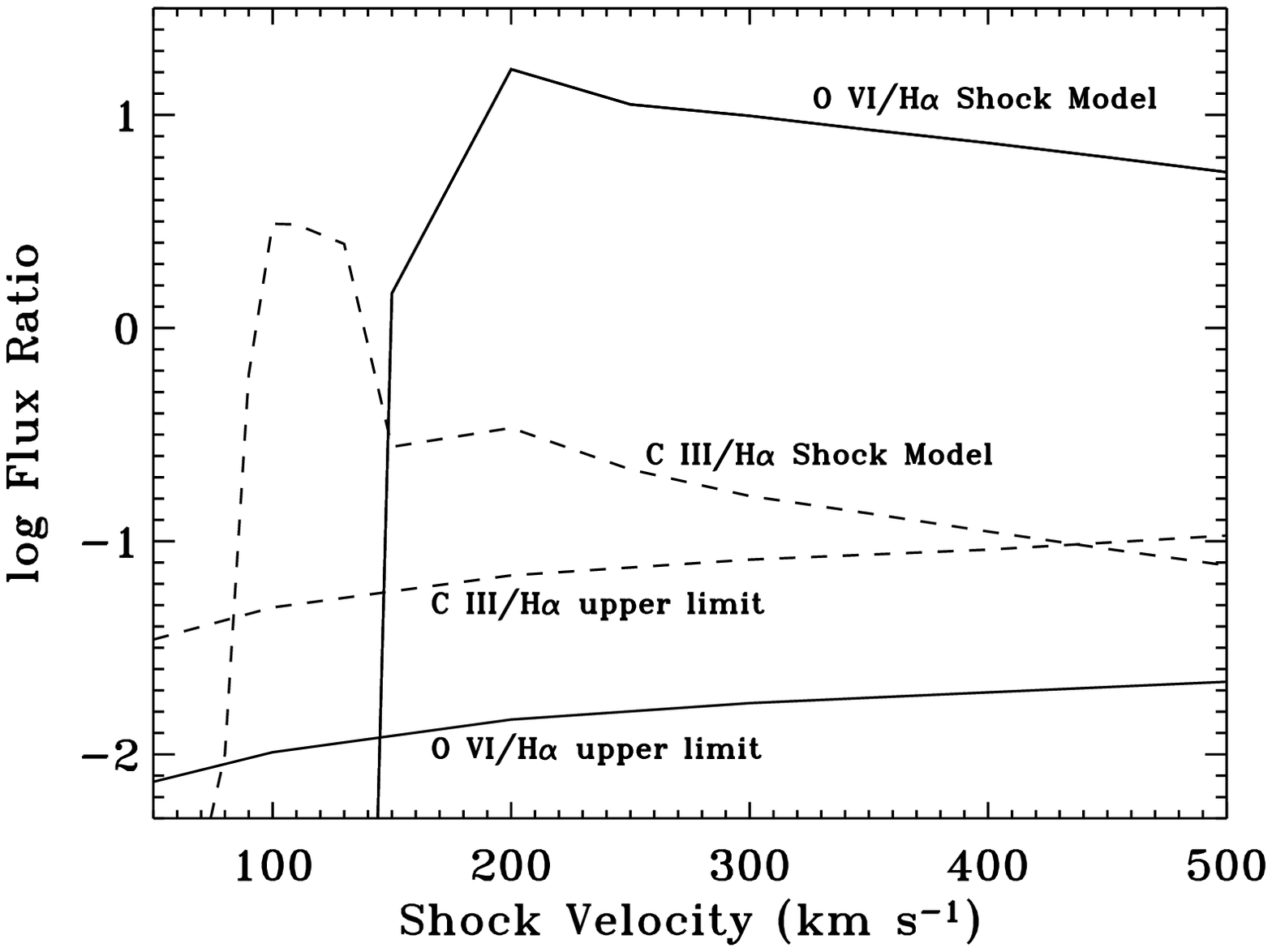}
\caption{Shock model predictions of O~VI/H$\alpha$ (solid line) and
C~III/H$\alpha$ (dashed line) from the models \cite{ds96}
(v$_s\ge150$~\kms) and \cite{sm79} (v$_s<150$~\kms). The nearly
diagonal lines show the observed upper limits at position A, the most
restrictive limit of the four pointings. The upper limits are
calculated for a line width (FWHM) corresponding to the shock velocity
(assuming bulk motion, not thermal broadening, is responsible for the
line width). The upper limits are well below the predictions,
indicating that most of the H$\alpha$-emitting gas is not shock
heated, unless the shocks have $v_s\la90$~km~s$^{-1}$.}
\end{figure}

Comparing the \ion{O}{6} and \ion{C}{3} limits with the \ha\ flux can
also constrain the conditions in the superwind. If a turbulent mixing
layer between hot and cold gas forms in the wind, the intermediate
temperature gas produced must be cooler than $10^{5.5}$~K, or
F$_{O~VI}$/F$_{H\alpha}$ would be higher than the observed limits in
Table 2 \citep{ssb93}. Figure 3 shows the predictions of shock models
for the \ion{O}{6}/\ha\ and \ion{C}{3}/\ha\ flux ratios
\citep{sm79,ds96}, compared to the observed upper limits on the ratios
at position A. The upper limits shown were calculated assuming a line
width (FWHM) corresponding to the shock velocity, so the limits are
higher for broader lines. The limits for all four pointings are listed
in Table 2, assuming the line width is 100~\kms.  The
F$_{O~VI}$/F$_{H\alpha}$ upper limit is lower than the prediction for
shocks with $v_s\ga150$~km~s$^{-1}$, and the F$_{C~III}$/F$_{H\alpha}$
ratio is lower than the prediction for shocks with
${v_s\ga90}$~km~s$^{-1}$, assuming all of the \ha\ emission arises in
shock-heated gas (very fast shocks are allowed by the
F$_{C~III}$/F$_{H\alpha}$ upper limit, but this is ruled out by the
F$_{O~VI}$/F$_{H\alpha}$ upper limit).  If the model of \cite{l99} is
correct, the slow shocks in the cold clouds that produce \ha\ emission
have ${v_s\la90}$~km~s$^{-1}$, or the ionization of the cold clouds is
dominated by photoionization from the starburst. CLOUDY models
\citep{ferland98} show that the F$_{C~III}$/F$_{H\alpha}$ limit is
consistent with photoionization (this will be discussed further in a
future paper, C. Hoopes et al., in preparation). If the clouds are
photoionized, this means that the energy source powering the \ha\
emission is not the wind, so even the optical emission does not
represent energy lost from the wind fluid.  This scenario raises the
possibility that the wind blown cavity provides a conduit for ionizing
photons from the starburst to reach the IGM.

\acknowledgments
We thank Ken Sembach for helpful discussions. We appreciate the
helpful comments from the anonymous referee. This work was supported
by NASA grant NAG5-11945.

\end{document}